\documentclass[prb,twocolumn,showpacs,preprintnumbers,amsmath,amssymb,superscriptaddress]{revtex4-1}

\usepackage{dcolumn}
\usepackage{bm}
\usepackage{amsmath,graphicx}
\usepackage{color}

\usepackage{bm}
\usepackage{url}

\begin{document}
\title{Corrugation of an unpaved road surface under vehicle weight}

\author{Chiharu Matsuyama}
\affiliation{Department of Environmental Sciences, University of Yamanashi, 4-4-37, Takeda, Kofu, Yamanashi 400-8510, Japan}

\author{Yukihiro Tanaka}
\affiliation{Division of Applied Physics,
Faculty of Engineering, Hokkaido University, Kita 13 Nishi 8, Sapporo 060-8628, Japan}

\author{Motohiro Sato}
\affiliation{Division of Mechanical and Aerospace Engineering,
Faculty of Engineering, Hokkaido University, Kita 13 Nishi 8, Sapporo 060-8628, Japan}

\author{Hiroyuki Shima}
\email{hshima@yamanashi.ac.jp}
\thanks{(Correspondence author)} 
\affiliation{Department of Environmental Sciences, University of Yamanashi, 4-4-37, Takeda, Kofu, Yamanashi 400-8510, Japan}

\date{\today}

%***********
\begin{abstract}
Road corrugation refers to the formation of periodic, transverse ripples on unpaved road surfaces. It forms spontaneously on an initially flat surface under heavy traffic and can be considered to be a type of unstable growth phenomenon, possibly caused by the local volume contraction of the underlying soil due to a moving vehicle's weight. In the present work, we demonstrate a possible mechanism for road corrugation using experimental data of soil consolidation and numerical simulations. The results indicate that the vertical oscillation of moving vehicles, which is excited by the initial irregularities of the surface, plays a key role in the development of corrugation.
\end{abstract}
%***********

%% KEYWORD
%% HIGH-LIGHT 

% \end{frontmatter}
\maketitle

%**********************************************************
\section{Introduction}
\begin{figure}[ttt]
  \centering
  \includegraphics[width=0.48\textwidth]{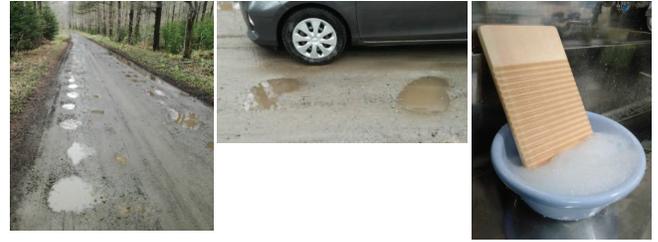}
\caption{Left: Road corrugation on unpaved road. Terrain undulations are clearly identified due to the rainwater collected in the depressions. Middle: Comparison of the period of corrugation with vehicle dimensions. Right: Photograph of a washboard.}
\label{fig_01}
\end{figure}

%**********************************************************
Globally, there are millions of kilometres of unpaved road (e.g., dirt roads, gravel roads). Such roads are less resilient than paved roads because their loose aggregate makes them vulnerable to gradual erosion. They thus deteriorate gradually from an initially flat state to having ruts, potholes, and other kinds of imperfections, which are formed by the repeated passage of vehicles, weather extremes, or both. Corrugation, a type of road deterioration, appears as many transverse ridges that form a periodic waveform along the vehicle travel direction. Corrugation develops spontaneously from an initially smooth surface. In a well-developed state, the amplitude and wavelength can be up to a few centimetres and several tens of centimetres, respectively \cite{Mather1963}. Typically, well-developed corrugation is likely to form on long flat unpaved roads, such as those found in South Dakota, U.S.A. \cite{Mahgoub2011}, and  the Outback in Australia \cite{Mather1963}. Corrugated roads are often called washboard roads (see Fig.~\ref{fig_01}).

Road corrugation causes vehicle vibration, which leads to vehicle occupant discomfort. Furthermore, it increases the risk of traffic accidents because it reduces the tyre-road surface contact area. The mitigation of road corrugation has thus long been a challenge for road maintenance \cite{Mahgoub2011,Alhasan2015}. The spontaneous formation of corrugation has attracted much academic attention because it seems counterintuitive. On first thought, downward compression due to the weight of moving vehicles should even out any initial road surface irregularities. In reality, however, roads subjected to more traffic more frequently develop surface corrugation. It is also interesting to note that similar corrugation behaviour can be observed when a fluid flows over rocks \cite{Veysey2008,Meakin2010,Jamtveit2012,Vesipa2015}, ice surfaces \cite{Camporeale2017,ASHChen2017}, or granular materials \cite{Zoueshtiagh2003} and when a rigid object (e.g., a plough) is pulled over a granular surface \cite{Bitbol2009,Hewitt2012,Percier2013,Srimahachota2017}. In addition, a corrugated bottom of a channel can induce physical anomalies in propagation of water waves along it \cite{Piat2012}. The mechanism of spontaneous surface corrugation is thus of research interest.

%----------------
%----------------

Experimental \cite{Mather1963,Stoddart1982,Taberlet2007,Bitbol2009} and numerical studies \cite{Mays2000,Both2001,Shoop2006,Taberlet2007,Ozaki2015,daSilva2018} have suggested that neither the shape and size of soil particles nor the underlying soil thickness significantly contribute to the development of road corrugation. The cohesive force between soil particles \cite{Mather1963} and the clay-sand ratio in soil \cite{Stoddart1982} have been suggested to be potential determinants of spontaneous surface corrugation. Furthermore, vehicle speed and tyre stiffness are possibly involved. In addition to these findings, it is believed that soil volume contraction likely plays a role in the corrugation mechanism. However, confirming this role experimentally is both laborious and time-consuming because many vehicles must travel over an unpaved road many times for the underlying soil thickness to contract. One promising alternative is to characterise the compressibility of soil samples using consolidation tests and then deduce the degree of volumetric shrinkage of the soil due to vehicle weight from the measured data. With measurement data of soil compressibility, vehicle-weight-induced corrugation phenomena can be numerically reproduced with high accuracy.

In the present study, we conducted soil consolidation tests to evaluate the characteristic quantities that govern the time-varying hardness of the soil under a compressive load. Using the measurement data, we performed numerical simulations of the time evolution of the shape and height of an unpaved road surface. The obtained results shed light on the role of soil contraction in the spontaneous formation of road corrugation.

%**********************************************************
\section{Consolidation experiment}
%**********************************************************

%-----------
\subsection{Soil samples}

Soil consolidation is a mechanical phenomenon in which soil subjected to continuous compressive force undergoes volume shrinkage with time. In this study, we measured the downward displacement and sinking rate of the top surface of a soil sample compressed downward over several hours. For the measurement, we used Keto soil, a kind of peat soil commercially available in Japan. This soil is soft and sticky when wet and very hard when dry. These characteristics are similar to those of actual unpaved road soil that exhibits spontaneous corrugation; the road surface is softened to some extent when wet and hardens as the soil dries.
In particular, wetting driven by spring thaw is known to cause significant softening of unpaved road even when formed of dry materials such as sand or gravel \cite{Shoop2006}. Therefore, our approach based on a moist soil sample may also be applicable to sand- or gravel-road corrugations during spring thaw.

%-----------
\subsection{Methodology}

\begin{figure}[ttt]
  \centering
\includegraphics[width=0.45\textwidth]{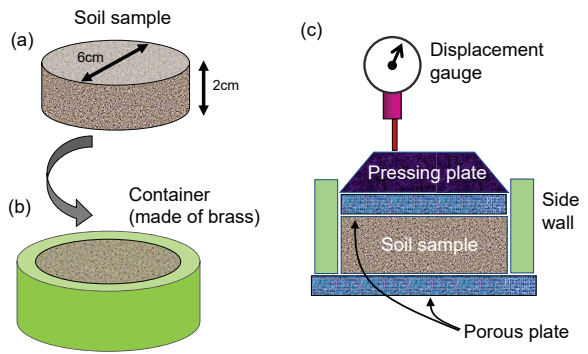}
\caption{Experimental procedure of soil consolidation. (a-b) After the soil sample is sufficiently stirred with a spatula, it is packed into a cylindrical brass container. (c) The container is set onto the consolidation test apparatus and a compressive load is applied to the top surface of the soil sample for a long time. During loading, water and air inside the soil seep out through the upper and lower porous plates.}
\label{fignew_02}
\end{figure}

Figure \ref{fignew_02} shows a diagram of the experimental method. First, we mixed a soil sample with a spatula until the water distribution inside the sample became uniform, and then packed it into a cylindrical container to prepare a specimen 6 cm in diameter and 2 cm in height. Next, we set the specimen in a consolidation apparatus and placed a weight with mass $M$ onto its top surface. The time variation of the top surface height, $H(t)=H_0+\delta H(t)$, was recorded at intervals for 6 hours. After 6 hours of compression, the specimen was replaced by a new pristine one with the same height and diameter as those of the previous specimen, and then a heavier weight was placed on the surface. The weight was sequentially changed from $M=0.4$ to 6.4 kg. The time intervals and weight used in the measurement are shown later (see Fig.~\ref{fignew_03}).

Using the measured data, we evaluated the softness coefficient of the soil, $c(\delta H, P)$, defined by
\begin{equation}
c(\delta H, P) = -\frac{1}{P}\cdot \frac{\partial H}{\partial t}.
\label{eq008}
\end{equation}
Here, $P=Mg/S_0$ is the downward pressure exerted on the top surface of the specimen, where $S_0= \pi r_0^2$, with $r_0=3$ cm and $g$ being the gravitational acceleration. Note that $c(\delta H, P)$ is positive for arbitrary $H$ and $P$ because $\partial H/\partial t<0$ under downward compression.

The softness coefficient $c(\delta H, P)$ is a proportional constant that relates the external pressure $P$ applied to the soil surface and the rate of plastic deformation of the soil $\partial H/\partial t$. It thus quantifies the magnitude of soil softness. A large (small) value of $c(\delta H, P)$ indicates that the soil is porous (dense); that is, the surface will subside greatly (barely) under the given downward pressure $P$. We show later that $c(\delta H, P)$ is a key quantity for simulating the effect of vehicle weight on the occurrence of unpaved road corrugation (see Eq.~\ref{eq002}).

It should be noted that the cylindrical brass sidewall shown in Fig.~\ref{fignew_02}(c) prevents lateral swelling of vertically loaded soil samples. In general, soft soil samples subjected to vertical load not only compress in vertical direction but also expand to some extent in lateral direction. The lateral displacement of a portion of the soil sample and consequent lateral drainage will promote the vertical settlement of the top surface of the sample, although this effect is not taken into account in the following discussion.

%----------------
%----------------

%-----------
\subsection{Soil volume contraction under weight}

\begin{figure}[ttt]
  \centering
\includegraphics[width=7.5cm]{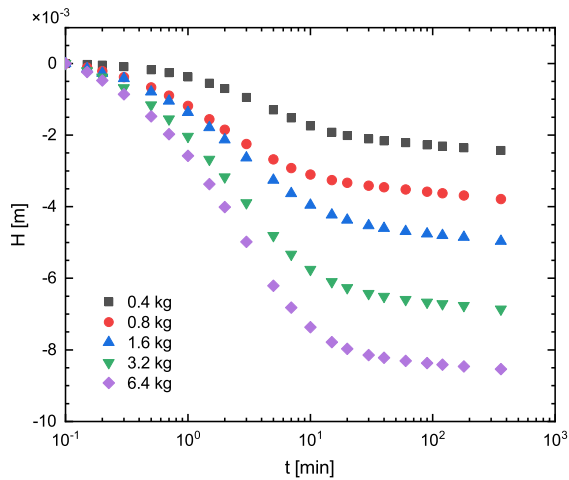}
\caption{Consolidation curve for 2-cm-thick soil samples subjected to various weights.}
\label{fignew_03}
\end{figure}

Figure \ref{fignew_03} shows the time variation of $H(t)$ obtained from the consolidation test. The horizontal axis shows the logarithm of elapsed time $t$ and the vertical axis shows the settlement of the surface from the initial height ($H_0=0$). The figure shows that as time passes, $H$ decreases monotonically for every weight condition. The value at which $H$ eventually converges depends on the weight. In addition, the settling speed of the surface (i.e., the slope of the $H$ curve) greatly changed at around 10 minutes of elapsed time.

\begin{figure}[ttt]
  \centering
\includegraphics[width=0.45\textwidth]{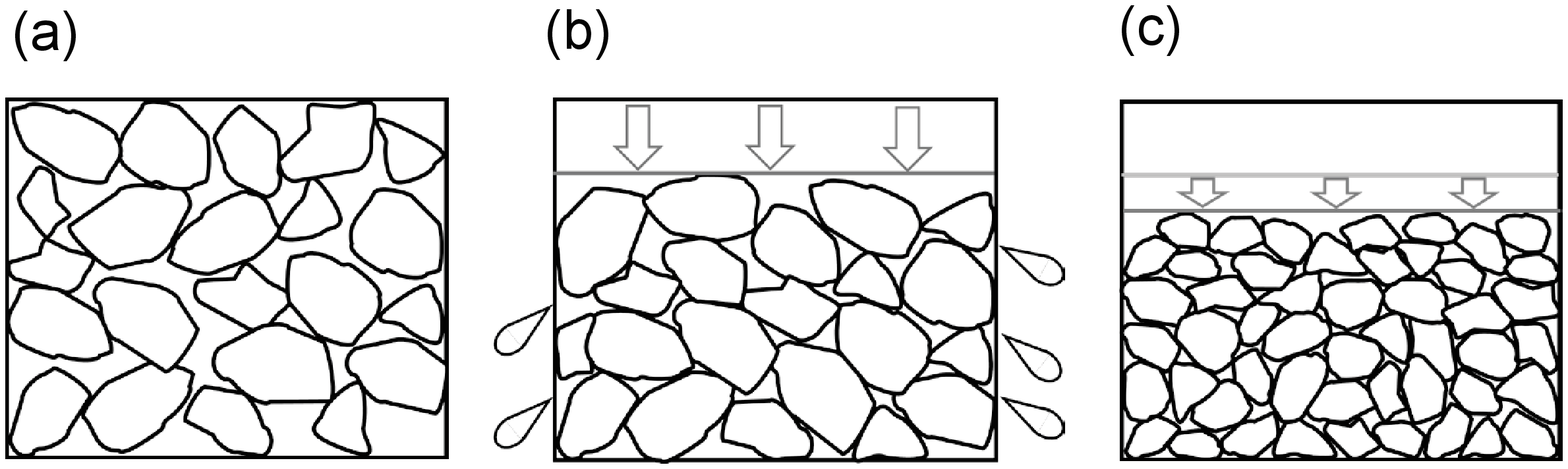}
\caption{Diagrams showing two consolidation processes. (a) Initial state. (b) Primary consolidation stage. (c) Secondary consolidation stage.}
\label{fignew_04}
\end{figure}

The transition in the slope of the $H(t)$ curves at around 10 minutes is attributed to the difference in the mechanism of soil volume contraction (see Fig.~\ref{fignew_04}). At $t <10$ min, volume contraction is mainly driven by the extrusion of pore water and pore air contained in the initial soil sample; this process is called primary consolidation [Fig.~\ref{fignew_04}(b)]. At $t> 10$ min, volume contraction continues after the air and water have been completely removed. This shrinkage is thought to result from a reconfiguration of the soil particle skeleton, where some soil particles collapse and become finer, filling the gaps in the original structure. This kind of soil settlement, caused by particle collapse and rearrangement, is called secondary consolidation [Fig.~\ref{fignew_04}(c)]. Because the finer soil particles are extremely small, the volume change through secondary consolidation is much smaller than that through primary consolidation.

%----------------
%----------------

%----------------
%----------------

%----------------------------------
\subsection{Softness coefficient}

\begin{figure}[ttt]
  \centering
\includegraphics[width=7.5cm]{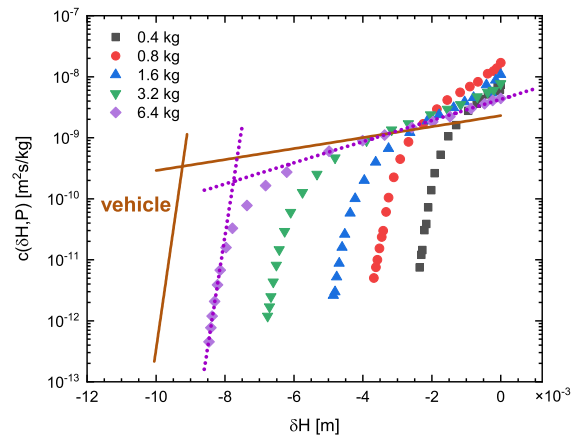}
\caption{Softness coefficient $c(\delta H, P)$ vs. displacement of the top surface of the soil sample $\delta H$. Dotted lines are two asymptotes for the 6.4-kg data with $\delta H_c \simeq -8$ mm, and solid lines show presumed asymptotes of $c(\delta H, P)$ for real-vehicle weight ($M=1000$ kg).}
\label{fignew_05}
\end{figure}

Figure \ref{fignew_05} shows a semilog plot of $c(\delta H, P)$ vs. $\delta H$ computed by substituting the measurement data of $H(t)$ into Eq.~(\ref{eq008}). As shown, $c(\delta H, P)$ decreases as the top surface of the specimen is depressed; in other words, the soil becomes gradually harder as sedimentation progresses. It is also understood from Fig.~\ref{fignew_05} that the value of $c(\delta H, P)$ depends not only on $\delta H$ but also on the downward pressure $P$.

%----------------
%----------------

A noteworthy finding from Fig.~\ref{fignew_05} is a crossover in the slope of the $c(\delta H, P)$ curve across a certain critical displacement $\delta H_c$. In the data for $M=6.4$ kg, for example, a considerable difference is found in the slope of the two asymptotic lines to the $c$ curve at $\delta H>\delta H_c$ and $\delta H < \delta H_c$ with $\delta H_c \simeq -8$ mm, both depicted by dotted lines in Fig.~\ref{fignew_05}. The crossover behaviour in the slope of $c$ curves is observed for every magnitude of weight. The value of $\delta H_c$ shifts monotonically from right to left with increasing weight.

The sudden change in the slope across $\delta H_c$ is a manifestation of 
the difference in the volume contraction mechanism mentioned above. 
That is, in Fig.~\ref{fignew_05}, the initial time regime 
(on the right side of $\delta H_c$), in which the slope of the $c$ curve is 
gentle, corresponds to the primary consolidation, 
and the latter time regime (on the left side of $\delta H_c$), in which the 
slope is steep, corresponds to the secondary consolidation. 
In the primary consolidation, the gaps between soil particles are not filled much
through the escape of pore air and pore water. 
Therefore, the hardness of the whole soil does not increase very much, 
as shown in Fig.~\ref{fignew_05}. 
In the secondary consolidation, 
some soil particles collapse and the interstices are 
filled with these fine particles, and thus the hardness of the whole soil increases rapidly. 
This yields the steep slope of the $c$ curve in the secondary consolidation region.

As a supplementary experiment, we conducted the consolidation test using volcanic ash soil, which is much drier than Keto soil in the initial state. We then confirmed that there is still a primary consolidation stage in which $\log c$ decreases linearly with decreasing $\delta H$, similar to what we observed in Fig.~\ref{fignew_05}. But for dry soil, the slope of $\log(c)$-line was much smaller than for wet soil, so the system was unable to reach the secondary consolidation stage within our experimental period. These facts imply that the proportionality relation between $\log(c)$ and $\delta H$ holds true, at least in the initial time regime, regardless of the moisture content of the soil sample at the initial stage.

%----------------------------------
\subsection{Vehicle-weighted $c$ curve estimation}

As can be seen in Fig.~\ref{fignew_05}, the threshold value $\delta H_c$ and the slopes of the asymptotes
change systematically with increasing weight $M$ (i.e., pressure $P$). 
Using this result, the value of $c(\delta H, P)$ for an arbitrary set of values $H$ and $P$,
including those in a real-vehicle situation,
can be estimated through the following procedure.

\begin{figure}[ttt]
  \centering
\includegraphics[width=0.45\textwidth]{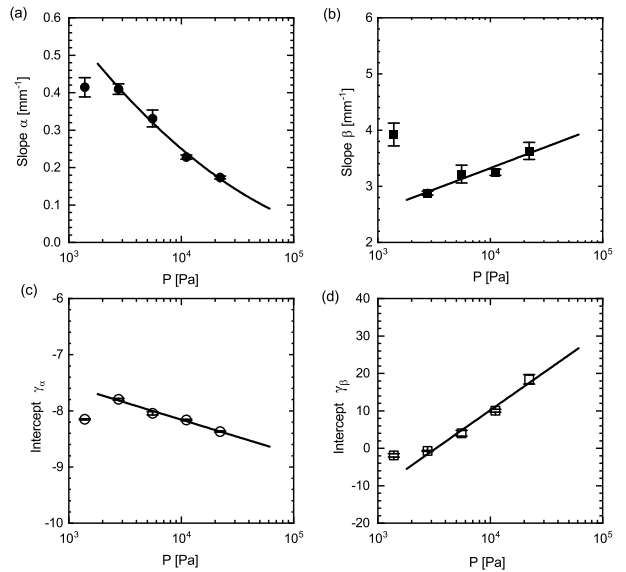}
\caption{Optimal values of $P$-dependent parameters (a) $\alpha$, (b) $\beta$, (c) $\gamma_\alpha$, and (d) $\gamma_\beta$. The solid curve in each plot is a quadratic fitting curve obtained using the least-squares method with the leftmost outlier point excluded.}
\label{fignew_06}
\end{figure}

We approximate the smooth $c$ curves by a combination of two asymptotic lines as below.
\begin{equation}
\log c(\delta H, P) = \left\{ \begin{array}{ll}
  \alpha(P)\cdot \delta H + \gamma_\alpha(P) & {\rm at} \;\; \delta H> \delta H_c(P), \\
  \beta(P)\cdot \delta H + \gamma_\beta(P) & {\rm at} \;\; \delta H < \delta H_c(P).
  \end{array} \right.
\label{eq012}
\end{equation}
The optimal values of $\alpha$ and $\gamma_\alpha$ for a given $P$ are
deduced from the linear regression of the ten rightmost data points in Fig.~\ref{fignew_05}
(e.g., $-4.98$ mm $\le \delta H \le$ 0 mm for $M=6.4$ kg).
Similarly, those of $\beta$ and $\gamma_\beta$ are deduced from the linear regression
of the six leftmost data points 
(e.g., $-8.46$ mm $\le \delta H \le$ $-8.14$ mm for $M=6.4$ kg).
The optimal values obtained show monotonic $P$ dependences, as summarised 
in Figs.~\ref{fignew_06}(a)-\ref{fignew_06}(d).
In each plot, the data points are fitted with a quadratic curve.

The quadratic fitting curves depicted in Figs.~\ref{fignew_06}(a)-\ref{fignew_06}(d)
allow us to infer the $\delta H$ dependence of the $c(\delta H, P)$ curve 
at any value of $P$, including that corresponding to a vehicle weight.
If the weight of actual four-wheeled vehicles is assumed to be 1000 kg and 
the contact area per pneumatic tyre is set to be $0.04$ m$^2$ 
(i.e., 20 cm $\times$ 20 cm),
the downward pressure exerted by each pneumatic tyre on the ground is equal to
$6.13 \times 10^4$ Pa.
To apply the equivalent pressure to the specimen 
(contained in a cylinder 6 cm in diameter) 
used in the consolidation test, a weight of 17.7 kg should be loaded. 
Under these conditions, the parameters $\alpha$, $\gamma_\alpha$, $\beta$, and $\gamma_\beta$
are expected to take the values listed in Table \ref{table_01},
as deduced from the fitting curves given in Figs.~\ref{fignew_06}(a)-\ref{fignew_06}(d).
Using these values, the asymptotes of the $c$ curves for the given vehicle weight ($M=1000$ kg)
can be estimated; the results are depicted in Fig.~\ref{fignew_05} by dashed lines.
These asymptotes are used in the numerical simulations discussed in the next section.

%-------------------
\begin{table}[bbb]
  \centering
  \caption{Estimated parameter values used to determine the approximate curve of $c(\delta H, P)$ for four-wheeled vehicles with $M=1000$ kg.}
  \begin{tabular}{|c|c|c|} \hline
    \quad Parameter \quad & Estimate & Unit \\ \hline 
    $\alpha$ & $9.00\times 10^{-2}$ &  \quad mm$^{-1}$ \quad \\
    $\beta$ & \quad $3.92$ \quad & \quad mm$^{-1}$ \quad \\
    $\gamma_\alpha$ & $-8.64$ & n/a \\
    $\gamma_\beta$  & 26.7 & n/a \\
    $\delta H_c$  & $-9.23$ & mm \quad \\ \hline
  \end{tabular}
\label{table_01}
\end{table}
%-------------------

%----------------
%----------------

%**********************************************************
\section{Numerical simulations}
%**********************************************************

%----------------
\begin{figure}[ttt]
  \centering
\includegraphics[width=0.45\textwidth]{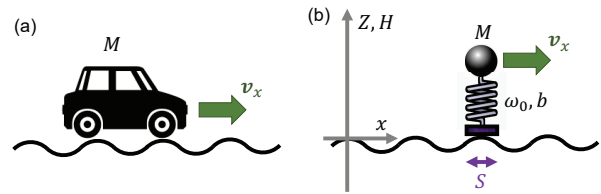}
\caption{(a) Sketch of an $M$-mass vehicle moving with velocity $v_x$. (b) Spring-mass model representing the vehicle moving in the $x$ direction. $Z(x,t)$ indicates the vertical position of the mass, $H(x,t)$ is the road surface height, and $S$ is the contact area. The point of reference for $Z(x,t)$ (i.e., $Z=0$) is defined so that $H(x,t)-Z(x,t)$ is the amount by which the effective spring is compressed.}
\label{fignew_07}
\end{figure}
%----------------

%-----------------------
\subsection{Effective spring model}

Vehicles traveling on an unpaved road 
exhibit vertical vibration during horizontal movement. Due to this vibration, which
is a mechanical response to surface irregularities,
impulsive downward forces are exerted intermittently
on the tyre-ground contact region;
at the same time, impulsive upward forces are applied to the pneumatic tyre at the point of contact.
If the temporal period of the impulsive forces is synchronised to that of the vertical vibration, 
the downward pressure caused by the vehicle weight is limited to 
specific regions that are equally spaced along the travel direction.
The accumulation of pressure in these regions may promote the development of road corrugation.
To examine the validity of this scenario,
we performed numerical simulations based on an extended version of
the theoretical model proposed in Refs.~\cite{Both2001,Kurtze2001}.

The primary assumption of the model is that the vertical vibration of vehicles
is equivalent to that of an effective damped harmonic 
oscillator, as earlier suggested by Ref.~\cite{Riley1973}. Hereafter, it is
referred to as the effective spring (see Fig.~\ref{fignew_07}).
In reality, vehicle vibration is determined by the mechanical properties of 
the suspension arm and shock absorber attached to the underbody
as well as those of pneumatic tyres, and is thus significantly more complicated
than that of a simple harmonic oscillator.
Nevertheless, in the present work,
we neglect the detailed structure of a real-vehicle underbody; 
instead, we aim to determine whether synchronization between
the temporal oscillation of vehicles and the spatial oscillation of the road surface
triggers the occurrence of road corrugation.
We also assume that the pneumatic tyres 
never lose contact with the road surface during travel.

%-----------------------
\subsection{Formulation}

The travel direction of the vehicle is set to be positive along the $x$ axis. The road surface height at position $x$ and time $t$ is denoted by $H(x,t)=H_0(x)+\delta H(x,t)$, and the vertical position of the vehicle centre of gravity is denoted by $Z(x,t)$; see Fig.~\ref{fignew_07}. The zero level of $H(x,t)$ is defined by the height of the initially flat road surface. Small irregularities are introduced at the start of the simulations. The zero level of $Z(x,t)$ is chosen such that $H(x,t)-Z(x,t)$ is  the amount by which the effective spring is compressed; i.e., the effective spring is compressed when $H>Z$ and stretched vertically when $H<Z$.

The equations of motion with these two variables are given by 
\begin{equation}
\frac{\partial H}{\partial t} = - c(\delta H, P)\cdot 
\left\| P+ \frac{M\omega_0^2 (H-Z)}{S} \right\|,
\label{eq002}
\end{equation}
\begin{equation}
M \frac{{\cal D}^2 Z}{{\cal D} t^2}
+
b \frac{{\cal D} (Z-H)}{{\cal D} t}
+
M \omega_0^2 (Z-H) = 0,
\label{eq004}
\end{equation}
with the notations
\begin{equation}
\| A \| = \left\{ \begin{array}{ll}
  A & {\rm when} \;\; A\ge 0, \\
  0 & {\rm when} \;\; A < 0,
  \end{array} \right.
\label{eq004a}
\end{equation}
and
\begin{equation}
\frac{{\cal D}}{{\cal D} t} = \frac{\partial}{\partial t} + v_x \frac{\partial}{\partial x}.
\label{eq006}
\end{equation}

Equation (\ref{eq002}) governs the time variation of the road surface height $H$. The quantity within the double vertical bars $\|\cdots\|$ is the downward compression term, which includes the eigenfrequency of the effective spring $\omega_0$. The notation defined by Eq.~(\ref{eq004a}) indicates that the tyres never pull upward from the ground at the point of contact. $c(\delta H, P)$ is the softness coefficient that we evaluated experimentally. For the vehicle-weight case, we substitute it with the values of the presumed asymptotes depicted in Fig.~\ref{fignew_05}.

Equation (\ref{eq004}) describes the vertical damped vibration of vehicles moving in the horizontal direction; $v_x$ is the horizontal velocity of the vehicle and $b$ characterises the magnitude of damping caused by energy dissipation in the underbody of the vehicle.

In principle, the second argument of the function $c(\delta H, P)$ should be not simply $P$ but equal to $P+M\omega_0^2 (H-Z)/S$, namely the total downward compression. In the present work, for simplicity, it is approximated simply by $P$, the downward compression in the static situation, considering the fact that the additional term,  $M\omega_0^2 (H-Z)/S$, oscillates (with sign changes) much faster than the characteristic time duration for the temporal change in $H$.

%----------------------------------
\subsection{Numerical conditions}

In the simulations, we set the numerical parameters as $M = 1000$ kg, $v_x = 10$ m/s ($= 36$ km/h), $S=0.16$ m$^2$, and $\omega_0=100$ rad/s (=15.9 Hz) assuming actual vehicle and road conditions; the validity of the $\omega_0$ value is examined later. The horizontal motion of the vehicle was limited to the range of $0\le x \le \lambda$ and the periodic boundary condition was applied to the $x$-direction. The time evolution computation was based on the MacCormack method \cite{cfdBook}, a popular finite-difference method for solving hyperbolic differential equations that is second-order-accurate in both space and time \cite{Mac1982,Mac2003}. The outline of the algorithm is given in the Appendix.

Before obtaining the time evolutions of $Z$ and $H$, initial imperfections with a sinusoidal form with wavelength $\lambda$ were introduced to $H_0$ as 
\begin{equation}
H_0(x) = \hat{h}_0 \sin \left( \frac{2\pi}{\lambda} x \right),
\label{eq024}
\end{equation}
with $\hat{h}_0 = 1.0$ cm.
The soil in the initial state was assumed to be uniformly non-consolidated. We then calculated the time-dependent growth of corrugation amplitude ${\cal G}(t)$, defined by
\begin{equation}
{\cal G}(t) = \Bigl| {\rm min} \left[H(x,t)\right] - {\rm max}\left[H(x,t)\right] \Bigr|.
\label{eq026}
\end{equation}
where ${\rm min}\left[ H(x,t) \right]$ and ${\rm max}\left[H(x,t)\right]$ are the minimum and maximum values of the road surface height, respectively, at a given time $t$ over the whole range of $x$.

We examine the time variation in ${\cal G}(t)$ by tuning the values of damping constant $b$ and wavelength $\lambda$, such that the numerical conditions of $b$ and $\lambda$ for the vertical amplitude of the initial imperfections grow with time. If ${\cal G}(t)$ increases monotonically during the whole elapsed time ($\sim$ 1 hour) in the simulation, the system is in the corrugated phase, in which the amplitude grows with time and thus well-developed corrugation will be eventually obtained. In contrast, if ${\cal G}(t)$ decreases monotonically, the system is in the flattened phase, in which the amplitude decreases with time. If ${\cal G}(t)$ oscillates or shows certain non-monotonic behaviour, the system is in the marginal phase.

%----------------------------------
\subsection{Results of numerical simulations}

\begin{figure}[ttt]
  \centering
\includegraphics[width=6.5cm]{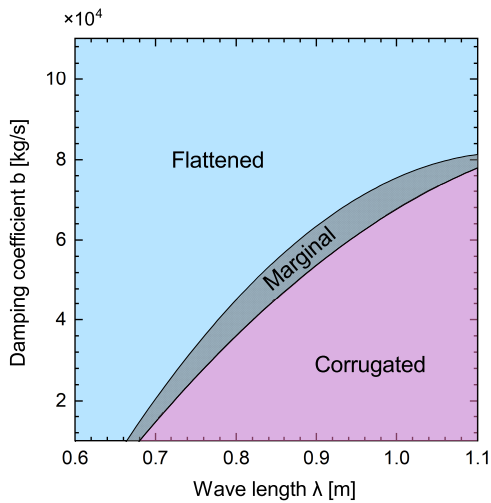}
\caption{
Phase diagram of corrugation occurrence in the $\lambda$-$b$ space.}
\label{fignew_08}
\end{figure}

Figure \ref{fignew_08} shows the phase diagram in the $\lambda-b$ space, illustrating the conditions of $\lambda$ and $b$ required for corrugation to occur under the present numerical conditions. In the corrugated phase (coloured red), a monotonic increase in ${\cal G}(t)$ with $t$ is confirmed, indicating the development of corrugation. In contrast, in the flattened phase (coloured blue), ${\cal G}(t)$ decreases monotonically with time and thus the initial sinusoidal imperfections level off. As shown in Fig.~\ref{fignew_08}, for a fixed $\lambda$ (and $\omega_0$), a small value of $b$ is preferred for corrugation to occur.

%----------------
%----------------

\begin{figure}[ttt]
  \centering
\includegraphics[width=0.45\textwidth]{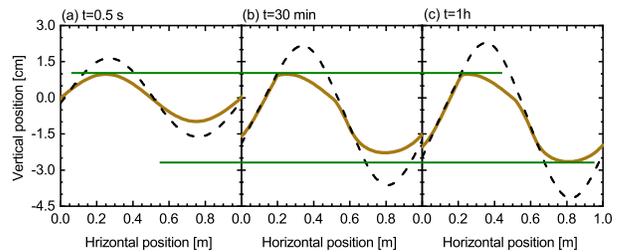}
\caption{
Time variation of the road surface profile $H(x)$ (solid curve) and the trajectory of the vehicle's centre of mass $Z(x)$ (dashed curve). The parameters are set as $\lambda=1.0$ m and $b=2.0 \times 10^4$ kg/s, corresponding to a corrugation-growing case. The two horizontal lines are provided as a visual guide for determining the degree of local sedimentation at the depression ($x\simeq 0.8$ m) and the persistence of the initial surface height at the bulge ($x\simeq 0.3$ m).}
\label{fignew_09}
\end{figure}

\begin{figure}[ttt]
  \centering
\includegraphics[width=0.45\textwidth]{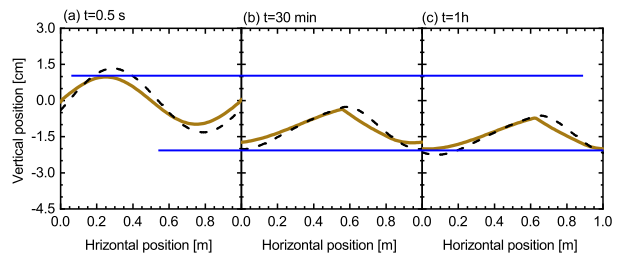}
\caption{
Time variation of $H(x)$ (solid curve) and the trajectory $Z(x)$ (dashed curve) in a corrugation-suppressed case with the settings $\lambda=1.0$ m and $b=10.0 \times 10^4$ kg/s.}
\label{fignew_10}
\end{figure}

For better understanding,
the time evolutions of both $Z(x,t)$ and $H(x,t)$ are visualised in Figs.~\ref{fignew_09} and \ref{fignew_10}, which
show the spatial profiles of $Z(x)$ (dashed curve) 
and $H(x)$ (solid curve) at the indicated times $t$. 
Two situations are considered; Fig.~\ref{fignew_09} shows 
the profiles under the corrugation-growing condition with the parameter settings 
$\lambda=1.0$ m and $b=2.0 \times 10^4$ kg/s, 
and Fig.~\ref{fignew_10} shows the profiles under the corrugation-suppressing condition 
with $\lambda=1.0$ m and $b=10.0 \times 10^4$ kg/s.

Figure \ref{fignew_09} shows that corrugation development is a consequence of
local sedimentation at the initial slight depression at around $x= 0.8$ m.
Near the depression, the inequality $Z(x)<H(x)$ holds for the whole elapsed time,
meaning that it is permanently compressed downward by the effective spring.
In contrast, the initial slight bulge at around $x= 0.3$ m
is not very strongly compressed because $Z(x)>H(x)$ for the whole elapsed time.
When the spring is stretched vertically more than 1 cm,
as shown at $x \simeq 0.3$ m in Figs.~\ref{fignew_09}(b) and \ref{fignew_09}(c),
the term in the double vertical bars of Eq.~(\ref{eq002}), $P+ [M\omega_0^2 (H-Z)]/S$,
becomes negative and thus no compressive force is exerted on the ground.
As a result of this tyre biased distribution of downward pressure,
the difference in the surface height grows with time.
This growth proceeds until the local sedimentation at the depression
reaches the secondary consolidation stage,
at which the soil becomes too hard to undergo further downward contraction.

%----------------
%----------------

%----------------
%----------------

A contrasting argument to the above leads us to a plausible mechanism by which the surface undulation becomes suppressed with time. In the case of Fig.~\ref{fignew_10}, the waveforms of $Z(x)$ and $H(x)$ do not become synchronised, as opposed to the case of Fig.~\ref{fignew_09}; i.e., the two waveforms are not in phase along the $x$ axis. As a result, compression occurs almost everywhere, causing a reduction in the initial surface undulation.

%**********************************************************
\section{Discussion}
%**********************************************************

Our numerical simulations were based on the assumption that the eigenfrequency of the effective spring is $\omega_0 = 100$ rad/s (i.e., 15.9 Hz). The validity of this assumption is discussed below. Suppose that road corrugation with a wavelength of $\lambda=$ 0.5 -- 1.0 m forms through the repeated passage of vehicles moving at speeds of $v_x=$ 10 -- 15 m/s (i.e., 36 -- 48 km/h); these values are derived from field observations of real road corrugation. For a vehicle's vertical oscillation to harmonise with the corrugation, the relation $\lambda \simeq v_x (2\pi/\omega_0)$ is expected to hold; this implies that the oscillation mode $\omega_0 \simeq$ 60 -- 180 rad/s (i.e., 10 -- 30 Hz) is relevant for the growth of road corrugation. It was argued in Refs.~\cite{Both2001,Ozaki2015} that the source of this oscillation mode is associated with the elastic vibration of the pneumatic tyres. Indeed, the lowest eigenfrequency of the pneumatic tyre vibration is on the order of tens of hertz \cite{Matsubara2017}, which is fairly consistent with the value estimated above.

The phase diagram of Fig.~\ref{fignew_08} shows that when $\lambda$ and $\omega_0$ are fixed, the damping constant $b$ plays a key role in deciding whether the initial surface undulation will grow or be suppressed. This finding is partly explained by considering two characteristic length scales, namely corrugation wavelength $\lambda$ and $v_x \tau$, which is the travel distance during the relaxation time, $\tau=2M/b$, of the spring's damped oscillation. Under the setting $M=1000$ kg, the value of $b=2.0\times 10^4$ kg/s, for instance, gives $\tau = 0.1$ s and $v_x \tau = 1.0$ m. This result of $v_x \tau$ indicates that the spring vibration is not completely damped while the vehicle travels a distance of one wavelength of the corrugation (0.5 -- 1.0 m). As a consequence, the spring vibration is synchronised with the road surface undulation, promoting the development of corrugation, as demonstrated in Fig.~\ref{fignew_08}. In contrast, the value of $b=10.0 \times 10^4$ kg/s corresponds to $\tau = 0.02$ s and $v_x \tau = 0.2$ m, implying that the spring vibration almost disappears before the vehicle travels a distance of one wavelength. Hence, synchronization does not occur and the initial surface undulation is likely to be suppressed until it reaches a flattened state.

It should be pointed out that this study only considered the effect of vertical load on road surface deformation. In addition to the vertical load, it is considered that the horizontal force generated on the tire-road interface (the force parallel to the traveling direction of the vehicle) also contributes to the corrugation formation. Existing work based on the finite element analysis \cite{Shoop2006,Ozaki2015} made clear that the frictional tire-road interaction plays a key role in the corrugation formation; this is consistent with field observation that the corrugation commonly occurs on transitional areas of the road such as around corners and slopes, where the vehicles applies large horizontal forces on the road through acceleration, deceleration, and steering \cite{Shoop2006}. Vehicle speed is also considered to be another important factor, as evidenced by existing work based on the discrete element method \cite{Taberlet2007}, which allows to analyse discontinuous behavior of soil particles.

Finally, we estimate the time required for the initial undulation to grow into well-developed corrugation. In our numerical simulations, we assumed that the road surface is constantly in contact with the tyres. However, a real vehicle moves on the road at a certain high speed, and thus the time that a tyre is in contact with a certain section of the road surface is very short. Suppose, for instance, that the contact area between the tyre and the road surface is 20 cm $\times$ 20 cm. The time that a vehicle moving at 10 m/s takes to travel a distance of 20 cm is estimated to be 0.02 s. If ten vehicles pass over this road every hour, the net time duration of the tyre-road contact will be 0.2 s per hour (approximately 30 minutes per year). This means that a few tens of minutes is sufficient for corrugation development in the simulations, even though real road corrugation may take around one year to develop. The time duration needed for real corrugation development depends on various conditions related to traffic and the local environment; its precise estimation is beyond the scope of the present work. Another issue is the effect of hysteresis; in a real unpaved road, the soil undergoes a series of short compressive cycles with each vehicle passing, and thus the time variation of $H$ may show hysteresis during the loading-unloading cycles. The effect of this possible hysteresis on the time evolution of the corrugation is an interesting topic for future work.

%**********************************************************
\section{Conclusion}
%**********************************************************

We performed consolidation experiments and numerical simulations to explore the mechanism of unpaved road corrugation. Based on measurement data of soil consolidation, we derived the softness coefficient $c(\delta H, P)$, whose $H$ and $P$ dependences determine the local sinking of the unpaved road surface under downward compression. Using the data, we numerically demonstrated that initial imperfections on the surface can grow with time under certain conditions due to the vertical vibration of moving vehicles. Based on an order estimation of the relationship between the vibration frequency, damping constant, and corrugation wavelength, we concluded that the elastic deformation of pneumatic tyres plays a role in corrugation development.

%*******************************************************
\section*{Acknowledgments}
We would like to thank Prof.~Shunji Kanie for fruitful discussions and Ms.~Etsuko Mukawa for technical assistance.
This work was supported by JSPS KAKENHI Grants (grant numbers 18H03818, 18K04879, 19K03766, and 19H05359).
%*******************************************************

%-----------------
\appendix

%******************************************************
\section*{Appendix A: Non-dimensionalisation}
%******************************************************

For computational convenience, the governing equations for $Z(x,t)$ and $H(x,t)$, 
given by Eqs.~(\ref{eq002}) and (\ref{eq004}), respectively, 
are non-dimensionalised. 
The time scale is set to $1/\omega_0$, the horizontal length scale is set
to $v_x/\omega_0$, and the vertical length scale is set to $g/\omega_0^2$. 
Accordingly, the dimensionless form of the governing equations is
\begin{equation}
\frac{\partial H}{\partial t} = -\mu(\delta H,P) (1+H-Z),
\label{eq305}
\end{equation}
\begin{equation}
\left( \frac{\partial}{\partial t} + \frac{\partial}{\partial x} \right)^2 Z
+
2\Gamma \left( \frac{\partial}{\partial t} + \frac{\partial}{\partial x} \right) (Z-H)
+
(Z-H) = 0.
\label{eq310}
\end{equation}
with new dimensionless parameters defined by
\begin{eqnarray}
\mu(\delta H,P) = M \omega_0 c(\delta H, P),\;\;
\Gamma = \frac{b}{2M \omega_0}.
\label{eq022}
\end{eqnarray}

%******************************************************
\section*{Appendix B: Reduction of differentiation order}
%******************************************************

Here, we introduce the function $Y(x,t)$ to transform Eq.~(\ref{eq310})
into a first-order differential equation with respect to both $x$ and $t$.
We define $Y(x,t)$ by
\begin{equation}
Y = \left( \frac{\partial}{\partial t} + \frac{\partial}{\partial x} \right) Z,
\label{eq315}
\end{equation}
which implies that
\begin{equation}
\frac{\partial Z}{\partial t} = -\frac{\partial Z}{\partial x} + Y.
\label{eq320}
\end{equation}
In addition, it follows from Eqs.~(\ref{eq310}) and (\ref{eq315}) that
\begin{equation}
\left( \frac{\partial Y}{\partial t} + \frac{\partial Y}{\partial x} \right)
+
2\Gamma Y - 2\Gamma \left( \frac{\partial H}{\partial t} + \frac{\partial H}{\partial x} \right)
+ (Z-H) = 0.
\label{eq335}
\end{equation}
Eliminating the term $\partial H/\partial t$ from Eqs.~(\ref{eq305}) and (\ref{eq335}) yields
\begin{eqnarray}
\frac{\partial Y}{\partial t} 
&=& -\frac{\partial Y}{\partial x} - 2 \Gamma Y 
- 2\Gamma \mu(H,P) \left( 1+H-Z \right) \nonumber \\
&+& 2 \Gamma \frac{\partial H}{\partial x} - (Z-H).
\label{eq340}
\end{eqnarray}

Equations (\ref{eq305}), (\ref{eq320}), and (\ref{eq340})
are the key equations that we solve using the discretization procedure.
In every equation,
the time derivative is approximated using a forward difference as
\begin{equation}
\frac{\partial R}{\partial t} \simeq \frac{R^{(n+1)}-R^{(n)}}{\Delta t},
\label{eq338}
\end{equation}
and the spatial derivative is approximated using a central difference as
\begin{equation}
\frac{\partial R}{\partial x} \simeq \frac{R_{j+1}- R_{j-1}}{2 \Delta x},
\label{eq350}
\end{equation}
where $R=H$ or $Z$ or $Y$;
the notations $R^{(n)} \equiv R(x,t)$ at $t=t_n$
and $R_j \equiv R(x,t)$ at $x=x_j$
were used in the discretization above.
Eventually, we obtain three sets of difference equations:
\begin{equation}
H_j^{(n+1)} = H_j^{(n)} + \Delta t
\left[
-\mu(H_j^{(n)},P) \left( 1+H_j^{(n)}-Z_j^{(n)} \right)
\right],
\label{eq381}
\end{equation}
\begin{equation}
Z_j^{(n+1)} = Z_j^{(n)} + \Delta t
\left[ \frac{Z_{j+1}^{(n)}- Z_{j-1}^{(n)}}{2 \Delta x} + Y_j^{(n)} \right],
\label{eq382}
\end{equation}
and
\begin{eqnarray}
Y_j^{(n+1)} &=& Y_j^{(n)} + \Delta t
\left[
\frac{Y_{j+1}^{(n)}- Y_{j-1}^{(n)}}{2 \Delta x} \right. \nonumber \\ 
&-& 2 \Gamma Y_j^{(n)} - 2\Gamma \mu(H_j^{(n)},P) \left( 1+H_j^{(n)}-Z_j^{(n)} \right) \nonumber \\
&+& \left. \Gamma \frac{H_{j+1}^{(n)}- H_{j-1}^{(n)}}{\Delta x} - \left( Z_j^{(n)}-H_j^{(n)} \right)
\right].
\label{eq383}
\end{eqnarray}

In the actual computation, the time evolution of Eqs.~(\ref{eq381}-\ref{eq383}) was 
realised using Heun's method (also called modified Euler's method) 
to ensure stability.
When the differential equation to be solved is given by
\begin{eqnarray}
\frac{dR}{dt}=f\left(R\left(t\right),t\right),
\end{eqnarray}
the general formula describing Heun's method is
\begin{eqnarray}
R^{(n+1)} &=&
R^{(n)}\!+\!
\frac{\Delta t}{2}\left[ f\left( R^{(n)},n\Delta t\right) \right. \nonumber \\
&+& f\left( R^{*},({n+1})\Delta t\right) \Big],
\label{eq385}
\end{eqnarray}
where $R^{*}$ is a predictor obtained using Euler's method:
\begin{eqnarray}
R^{*}=R^{(n)}+\Delta t f\left(R^{(n)},n\Delta t\right).
\label{eq386}
\end{eqnarray}
Equations (\ref{eq385}-\ref{eq386}) are 
the numerical procedure of Heun's method, 
in which, 
when deriving the solution at the next time step, 
the predictor $R^{*}$ is approximated using Euler's method 
with first-order accuracy in $\Delta t$
and the solution (corrector) in the next time step is given by 
an average value with the predictor.
This procedure yields a solution, $R^{(n+1)}$,
with second-order accuracy in $\Delta t$.

%%%%%%%%%%%%%%%%%%%%%%%%%%%%%%%%%%%%%%%%%%
\bibliographystyle{apsrev4-1}
\bibliography{CMCorrugation}

%%%%%%%%%%%%%%%%%%%%%%%%%%%%%%%%%%%%%%%%%%
\end{document}